\documentclass[%
 aip,
 amsmath,amssymb,
 reprint,%
]{revtex4-1}

\usepackage{graphicx}
\usepackage{dcolumn}
\usepackage{bm}
\newcommand{\dd}{\mathrm{d}}
\usepackage{siunitx}

\usepackage[utf8]{inputenc}
\usepackage[T1]{fontenc}
\usepackage{mathptmx}
\usepackage{etoolbox}
\usepackage{xcolor}

\makeatletter
\def\@email#1#2{%
 \endgroup
 \patchcmd{\titleblock@produce}
 {\frontmatter@RRAPformat}
 {\frontmatter@RRAPformat{\produce@RRAP{*#1\href{mailto:#2}{#2}}}\frontmatter@RRAPformat}
 {}{}
}%
\makeatother
\begin{document}

\preprint{AIP/123-QED}

\title[Kinetic Simulations Verifying Reconnection Rates Measured in the Laboratory, Spanning the Ion-Coupled to Near Electron-Only Regimes]{Kinetic Simulations Verifying Reconnection Rates Measured in the Laboratory, Spanning the Ion-Coupled to Near Electron-Only Regimes}
\author{S. Greess}
\email{greess@wisc.edu.}
\author{J. Egedal}%
\affiliation{Department of Physics, University of Wisconsin - Madison, Madison, Wisconsin 53706, USA}%

\author{A. Stanier}
\affiliation{Los Alamos National Laboratory, Los Alamos, New Mexico 87545, USA}%

\author{J. Olson}%
\affiliation{Department of Physics, University of Wisconsin - Madison, Madison, Wisconsin 53706, USA}%

\author{W. Daughton}
\affiliation{Los Alamos National Laboratory, Los Alamos, New Mexico 87545, USA}%

\author{A. L\^{e}}
\affiliation{Los Alamos National Laboratory, Los Alamos, New Mexico 87545, USA}%

\author{A. Millet-Ayala}%
\affiliation{Department of Physics, University of Wisconsin - Madison, Madison, Wisconsin 53706, USA}%

\author{C. Kuchta}%
\affiliation{Department of Physics, University of Wisconsin - Madison, Madison, Wisconsin 53706, USA}%

\author{C.B. Forest}%
\affiliation{Department of Physics, University of Wisconsin - Madison, Madison, Wisconsin 53706, USA}%

\date{\today}

\begin{abstract}
The rate of reconnection characterizes how quickly flux and mass can move into and out of the reconnection region. In the Terrestrial Reconnection EXperiment (TREX), the rate at which antiparallel asymmetric reconnection occurs is modulated by the presence of a shock and a region of flux pileup in the high-density inflow. Simulations utilizing a generalized Harris-sheet geometry have tentatively shown agreement with TREX's measured reconnection rate scaling relative to system size, which is indicative of the transition from ion-coupled toward electron-only reconnection. Here we present simulations tailored to reproduce the specific TREX geometry, which confirm both the reconnection rate scale as well as the shock jump conditions previously characterized experimentally in TREX. The simulations also establish an interplay between the reconnection layer and the Alfv\'{e}nic expansions of the background plasma associated with the energization of the TREX drive coils; this interplay has not yet been experimentally observed.
\end{abstract}

\maketitle

\section{\label{sec:intro}Introduction}
Magnetic reconnection \cite{dungey:1953} is the process through which the topology of magnetic field lines change in the presence of a plasma, often resulting in an explosive release of magnetic energy. Well-known examples include solar flares \cite{priest:2000} and auroral substorms in the Earth's magnetosphere \cite{vasyliunas:1975}. Reconnection is studied both {\sl in situ} in the magnetosphere by satellites like NASA's MMS mission \cite{burch:2016b} and in laboratory experiments here on Earth. One such experiment is the Terrestrial Reconnection EXperiment (TREX) at the University of Wisconsin-Madison \cite{Forest2015}, which is operated specifically to reach parameter regimes of magnetic reconnection relevant to those of collisionless space plasmas \cite{olson:2016}.
Previous results from TREX have characterized the rate at which reconnection occurs in the experiment \cite{olson:2021}. The reconnection rate determines the speed at which plasma and magnetic field lines can move into and out of the reconnection region, effectively setting the timescale of the entire process \cite{parker:1957,zweibel:2016,ji:2022}. These TREX results showed the importance of magnetic flux pileup and the formation of a shock preceding the reconnection layer in maintaining force balance and setting the normalized reconnection rate. Furthermore, the experimental rate has a dependence on the size of the system relative to the ion scale; smaller scale size produces higher rates, indicative of the transition from ion-coupled toward electron-only reconnection\cite{olson:2021,phan:2018,liu:2017}. 

Similar to previous numerical simulations of the TREX geometry\cite{greess:2021}, in this letter we will apply fully kinetic simulations with the aim to confirm and reproduce the results of Ref.~\onlinecite{olson:2021}. This is part of an ongoing effort to synchronize data collection between the experimental and simulated TREX environments, using the VPIC code developed at Los Alamos National Laboratory\cite{greess:2021,bowers_2020,2018APS..DPPC11023D,bowers:2009}. After a brief introduction to the TREX experiment and the VPIC code, multiple simulations of TREX will be analyzed to verify that pressure balance exists across the reconnection shock front. Finally, further simulations of TREX will be evaluated to check if the reconnection rate results match the values measured in TREX and their dependence on normalized experimental system size.

\begin{figure}
\includegraphics{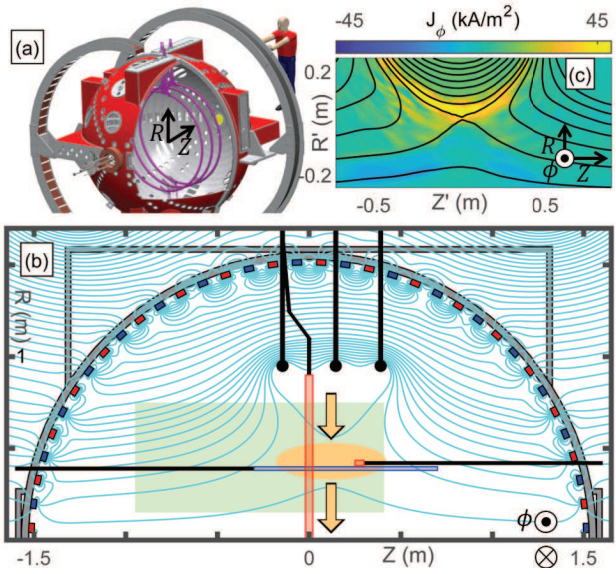}
\caption{(a) Engineering sketch of TREX. The internal drive coils (purple) drive a magnetic field that opposes the external Helmholtz coil's field. The plasma source is a polar array of plasma guns (yellow). (b) A cross-section of the top half of the TREX vessel showing a theoretical example of the typical experimental geometry. The magnetic field lines are shown in cyan. The reconnection region (light orange) is driven down from the drive coils to the central axis, as indicated by the arrows. The layer is measured during this transit by our diagnostic suite. (c) shows a plot of an experimental out-of-plane reconnection current layer, created by collating data taken from a single probe as it is moved through the green shaded region in (b). The black lines are contours of the flux function, $\Psi$, which map to the magnetic field lines. The data has been re-centered around the x-point.}
\label{fig:trex}
\end{figure}

\subsection{\label{sec:trex}The Terrestrial Reconnection EXperiment (TREX)}
An engineering schematic of TREX is presented in Fig.~\ref{fig:trex}(a). The vacuum vessel, provided by the Wisconsin Plasma Physics Laboratory (WiPPL)\cite{Forest2015}, is a $3~\si{\meter}$ diameter sphere that uses an array of permanent magnets embedded in the chamber wall to limit the plasma loss area to a very small fraction of the total surface area while keeping the bulk of the plasma unmagnetized. The setup includes a set of internal drive coils and an exterior Helmholtz coil that provides a near-uniform axial magnetic field with a magnitude up to $28~\si{\milli\tesla}$ \cite{olson:2016,olson:2021}. The current through the three internal drive coils (purple) ramps up to create a magnetic field that opposes and reconnects with the background Helmholtz field, resulting in an anti-parallel magnetic configuration (e.g. no significant guide field). The plasma source is a set of plasma guns located at the machine's pole (shown in yellow). This setup mimics the asymmetric conditions of the dayside magnetopause; the high-density, low-field inflow at low $R$ (analogous to the solar wind) is opposed by the low-density, high-field inflow at high $R$ (analogous to the Earth's magnetic field). TREX is typically operated in either hydrogen, deuterium, or helium plasmas.

In the planar cut of TREX shown in Fig.~\ref{fig:trex}(b), the cyan lines illustrate the typical magnetic geometry of an experimental run. As the current through the drive coils ramps up, the reconnection region is pushed from underneath the drive coils radially inward (orange arrows in Fig.~\ref{fig:trex}(b)). With a typical reconnection layer speed of $50~\si{\kilo\meter}/\si{\second}$, the temporal resolution of our probes ($10~\si{\mega\hertz}$) translates to a high spatial resolution measurement of about $5~\si{\milli\meter}$. This process, referred to as the "jogging method", permits the magnetic structure of the entire reconnection geometry to be characterized in a single experimental shot. The various magnetic and temperature probes and their locations are represented by the blue and red rectangles in Fig.~\ref{fig:trex}(b). In addition to the jogging method probes, a different array of 3-axis $\mathbf{\dot{B}}$ probes can be moved between shots, allowing for the creation of multi-shot datasets. The coverage area of this probe is given by the light green rectangle in Fig.~\ref{fig:trex}(b).

An example of data collected from a typical set of experimental shots is provided in Fig.~\ref{fig:trex}(c), where data from $34$ shots are combined into one picture; for each shot, the probe is at a different position within the green region in Fig.~\ref{fig:trex}(b). The black lines are contours of the flux function $\Psi$ to illustrate the in-plane magnetic field lines. Typical plasma parameters near the reconnection region include $T_i\ll T_e~\simeq 5 - 20 ~\si{\electronvolt}$, $n_e~\simeq 2\times10^{18} ~\si{\meter}^{-3}$, $B_{rec}~\simeq 4~\si{\milli\tesla}$, yielding $\beta_e~\simeq 0.4$ and $S~\simeq 10^4$. 

\subsection{\label{sec:vpic}Kinetic Simulation Model}
TREX is simulated using VPIC, a kinetic particle-in-cell code \cite{bowers:2009,2018APS..DPPC11023D,bowers_2020}. VPIC has previously been used to mimic the TREX setup and produce results comparable to experimental data. More information on the general usage of VPIC to simulate TREX may be found in Ref.~\onlinecite{greess:2021}.
The number of grid-points in the 2D simulations described here is $756$ by $1800$, spanning a system size of about $4.5$ by $9$ ion skin depths ($d_i$) in the $R$ and $Z$ directions, respectively, in our standard density case. The low $R$ boundary is set at $R\approx0.155d_i$ in the standard case and acts as a reflecting conductor; this is meant to replicate the effect of the cylindrical TREX current layer bouncing off of itself once it reaches $R=0~\si{\meter}$ in the experiment (cylindrical VPIC cannot operate at $R=0$, so a relatively close value is chosen for the lower bound instead; the closer this value is to $0$, the higher the computational load). The average number of macro-particles per cell is $500$. In all simulations presented here, the ratio of the electron cyclotron frequency to the electron plasma frequency is $1$ and the mass ratio $m_i/m_e$ is $400$. Sub-realistic mass ratios are typical of PIC simulations for computational tractability; as a consequence, the experimental size measured in $d_e$ is slightly larger than the simulation domain. However, Ref.~\onlinecite{olson:2021} ran the TREX experiment at different ion masses and verified that the reconnection rate is tied to $d_i$ rather than $d_e$. 
A typical experimental reconnection drive is modeled by injecting a linearly increasing current through the simulated drive coils. The strength of this current drive is given as a fraction of a typical experimental measurement of the time rate of change of the current in the drive coils, $\dot{I}_0\approx6.28\times10^{8}~\si{\ampere\per\second}$.

\section{\label{sec:pbalance}Regions of Reconnection and Pressure Balance}
\begin{figure}
\includegraphics{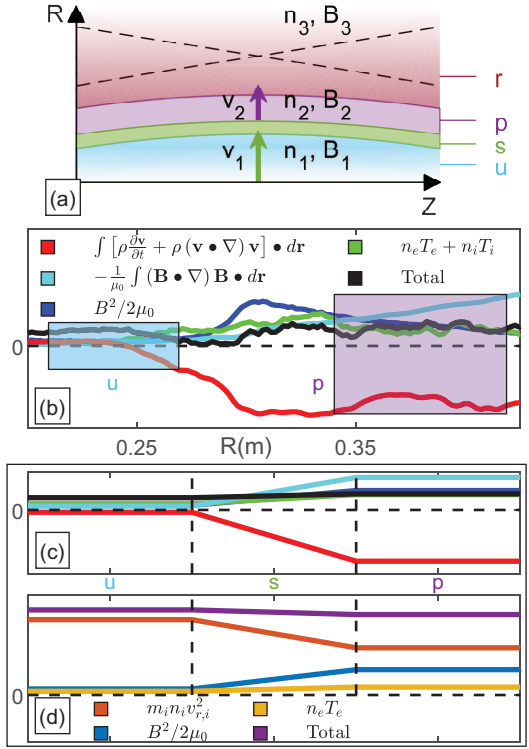}
\caption{(a) Cartoon showing the regions that compose the area below a reconnection layer in the experiment, as described in Ref.~\onlinecite{olson:2021}. At the lowest $R$ values, we start in region $u$, below the shock ($s$) that precedes the reconnection layer. Following the shock, the pileup region ($p$) comes before the reconnection region ($r$). All labeled velocities are shown to be in the reference frame of the shock layer ($s$). (b) A plot of how the different terms of the momentum/pressure balance equation in simulation data vary across the shock. The net momentum remains roughly constant moving from region $u$ (blue highlight) through the shock up to region $p$ (purple highlight). (c) The same data as subplot (b), but averaged over the highlighted regions to give single values for each term in regions $u$ and $p$. (d) A recreation of the momentum/pressure balance demonstration from Ref.~\onlinecite{olson:2021}. This analysis of simulation data applies the assumptions that were necessary for the initial experimental data evaluation in Ref.~\onlinecite{olson:2021}, which are borne out by the fact that the pressure is still balanced in this application of the method to simulation data. Subplots (b)-(d) are all in arbitrary code units. Subplot (a) reproduced from Olson, J., Egedal, J., Clark, M., Endrizzi, D., Greess, S., Millet-Ayala, A., and Forest, C., Journal of Plasma Physics, 87(3), 175870301, doi:10.1017/S0022377821000659, 2021, licensed under a Creative Commons Attribution (CC BY) license.}
\label{fig:pbalance}
\end{figure}

In TREX and TREX simulations during a reconnection discharge, the upstream plasma below the current layer can be divided into several distinct regions, as illustrated in Fig.~\ref{fig:pbalance}(a)\cite{olson:2021}. Working in the reference frame of the reconnection layer ($r$, red), the far upstream ($u$, blue) moves toward the layer at a speed faster than the local Alfv\'{e}n speed. This necessitates the formation of a region of magnetic flux pileup ($p$, purple); this region is separated from the far upstream by a sub-critical shock ($s$, green). Pressure balance between these regions was verified using TREX data in the shock's reference frame in Ref.~\onlinecite{olson:2021}; the assumptions and approximations involved with this calculation will be detailed below.

When combined with Amp\`{e}re's Law, the MHD momentum balance equation for the plasma in regions $u$ and $p$ is
\begin{equation}
 \rho \frac{\dd \mathbf{v}}{\dd t} = \frac{1}{\mu_0}\left(\boldsymbol\nabla \times \mathbf{B} \right) \times \mathbf{B} - \boldsymbol\nabla p\quad.
 \label{eq:mombal1}
\end{equation}
where $\rho$ is the plasma density, $\mathbf{v}$ is the plasma flow speed, $\mathbf{B}$ is the magnetic field, $p$ is the plasma kinetic pressure, and $\dd/\dd t$ is the total convective derivative, $\dd/\dd t = \partial/\partial t + \mathbf{v} \bullet \boldsymbol\nabla$. By plugging in the full convective derivative and using a vector identity, Eq.~\ref{eq:mombal1} can be rewritten as 
\begin{equation}
 \boldsymbol\nabla \left(p + \frac{B^2}{2\mu_0}\right) = \frac{1}{\mu_0}\left(\mathbf{B} \bullet \boldsymbol\nabla \right)\mathbf{B} -\rho \frac{\partial \mathbf{v}}{\partial t} - \rho\left(\mathbf{v} \bullet \boldsymbol\nabla\right)\mathbf{v}\quad.
 \label{eq:mombal2}
\end{equation}
Due to the toroidal symmetry of our experiment and the periodic boundary conditions in $\phi$ in our simulation, we assume that $\partial a/\partial \phi = 0$ for any quantity $a$.
To evaluate the pressure balance between regions $u$ and $p$, we will integrate Eq.~\ref{eq:mombal2} over a path $\dd \mathbf{r}$ between arbitrary points $g$ and $h$, where these points share the same value of the $Z$ coordinate such that $\dd \mathbf{r} = \dd r \mathbf{\hat{r}}$. The resulting momentum/pressure balance relation is
\begin{eqnarray}
	\left[p + \frac{B^2}{2\mu_0}\right]_g - \left[p + \frac{B^2}{2\mu_0}\right]_h - \frac{1}{\mu_0}\int^g_h \left(\mathbf{B} \bullet\boldsymbol\nabla \right)\mathbf{B} \bullet \dd \mathbf{r} \nonumber\\ + \int^g_h \left[\rho \frac{\partial \mathbf{v}}{\partial t} + \rho\left(\mathbf{v} \bullet \boldsymbol\nabla\right)\mathbf{v}\right] \bullet \dd \mathbf{r} = 0\quad. \label{eq:mombal3}
\end{eqnarray}
By taking $g$ and $h$ to be in regions $u$ and $p$ respectively, we can evaluate the change in the different terms of this equation across the shock that separates the two regions; this analysis is shown in Fig.~\ref{fig:pbalance}(b). Here, the equation has been split into distinct terms, where the magnetic pressure is in blue, the magnetic curvature is in cyan, the total convective acceleration is in red, the total kinetic pressure (given as the sum of the ion and electron pressures) is in green, with the total (the sum of all the terms) given in black. All terms are evaluated in the reference frame of the shock. The value for each term can be averaged over the points highlighted in regions $u$ and $p$ to give a single value for each, resulting in Fig.~\ref{fig:pbalance}(c); in this plot, the values in $s$ are simply the difference between the $u$ and $p$ regions. This averaging was done to mimic the limitations of the TREX experiment: the speeds of different regions cannot be measured simultaneously, so only a single value for each term can be calculated in each of regions $u$ and $p$. Both $u$ and $p$ are taken to be outside the electron diffusion region, such that the electron contribution to the inertia term is neglected\cite{olson:2021}. As expected, the total momentum/pressure is constant across the shock layer.

When evaluating the pressure balance across the shock layer in the experiment, several approximations are needed to account for the limitations of data collection (including the region speed limitation detailed above). Most notably, the analysis in Ref.~\onlinecite{olson:2021} assumed that in regions $u$ and $p$, changes in the plasma's velocity with respect to time or spatial coordinate are both minor relative to the ram pressure term, $m_i n_i v^2_{r,i}$, and balanced by the change in the magnetic tension tension term. The experimental analysis also assumed that $T_e \gg T_i$. Both of these assumptions are tested in Fig.~\ref{fig:pbalance}(d), where the simulation data is used to recreate TREX's measurements. The total of the approximate pressure terms, shown in purple, is constant across the shock, as it is in the full momentum balance analysis detailed above and shown in Fig.~\ref{fig:pbalance}(b) and (c). From this, we conclude that the assumptions that went into Ref.~\onlinecite{olson:2021}'s are also consistent with the presented numerical results.

\section{\label{sec:rate}Reconnection Rate}
\begin{figure*}
\includegraphics{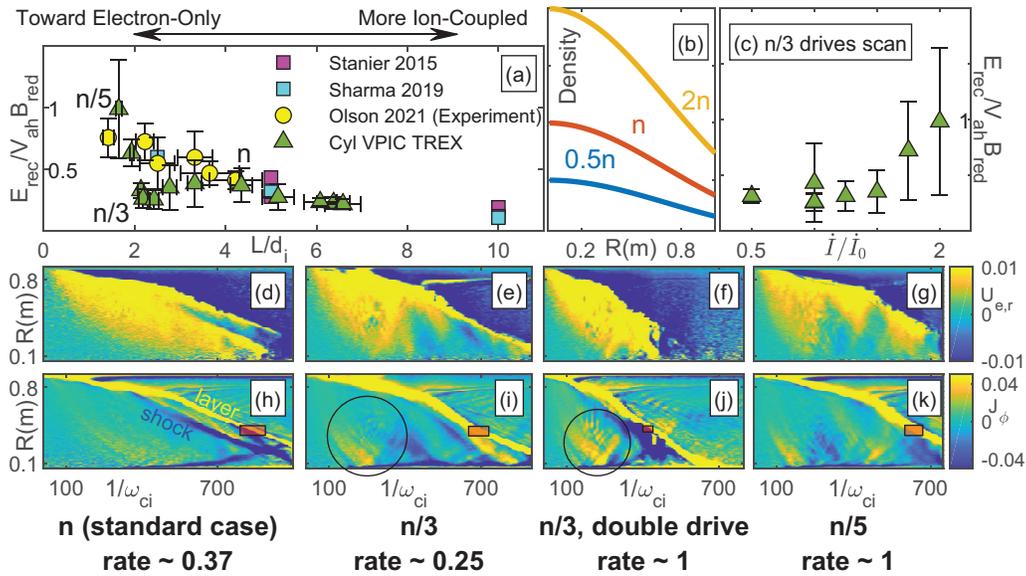}
\caption{(a) The rate results of our density scan simulations, compared with prior simulation results and the TREX experimental results from Ref.~\onlinecite{olson:2021}. The new rates (triangles) exhibit the same scaling as previous rate results, but with a slight dip around the $L/d_i\approx2$ region. (b) A rough demonstration of how the densities of our simulations were varied to control the scaled system size. Our initial density profile ($n$) is scaled up or down by a variety of factors from $1/5$ up to $4$. (c) Variation of the rate for density $n/3$, the bottom of the dip in the simulation points in (a). The three points that are averaged together to get the $n/3$ in (a) are each shown here at the $\dot{I}/\dot{I}_0$ $=1$ location. As the rate of current injection (ie, the strength of the drive) increases, dip feature disappears. (d)-(k) are profiles in the $R$ vs time plane (cuts of constant $Z$ through the x-point) of different variables for four different simulations, showing how the reconnection features change with the rate. The first column is our standard density case, the second is an $n/3$ scan (in the dip of (a)), the third is an $n/3$ scan with twice the standard drive (rightmost point in (c)), and the last is an $n/5$ scan. The first row (d)-(g) shows the radial electron velocity, $U_{e,r}$. The second row (h)-(k) shows the current layer density $J_{\phi}$; the layer and the preceding shock are labeled in the standard case (h) and visible with varying degrees of strength in the other cases. The black circles in (i) and (j) are showing the current due to the propagation and rebounding of the Alfv\'{e}nic perturbation. The red rectangles represent the location of the x-point when the reconnection rate is measured, corresponding to our experimental measurement region of $R=0.35-0.45~\si{\meter}$.}
\label{fig:rates}
\end{figure*}

As described previously, reconnection in TREX is asymmetric in plasma density and magnetic field on the opposing sides of the reconnection layer. As such, the reconnection rate is appropriately normalized by the method derived in Ref.~\onlinecite{cassak:2007}:
\begin{eqnarray}
 E_{rec} =& \alpha B_{red} V_{A,hyb} \quad, \label{eq:rate}\\
 B_{red} =& \frac{2B_2 B_3}{B_2 + B_3}\quad, \label{eq:bred}\\
 V_{a,hyb} =& \left[ \frac{1}{\mu_0 m_i}\frac{B_2 B_3 \left(B_2 + B_3\right)}{n_3B_2 +n_2B_3}\right]^{1/2}\quad. \label{eq:vah}
\end{eqnarray}
where $\alpha$ is the normalized reconnection rate, $E_{rec}$ is the reconnection electric field, $B_{red}$ is the reduced magnetic field, $V_{a,hyb}$ is the hybrid Alfv\'{e}n speed, and $B$ and $n$ are magnetic field and plasma density values at locations $2$ and $3$ as shown in Fig.~\ref{fig:pbalance}(a). Experimental values of $\alpha$ were calculated in TREX, where the reconnection electric field was derived from the time rate of change of the magnetic flux function, $\Psi$ \cite{olson:2021}. Results from this evaluation are shown as the yellow points in Fig.~\ref{fig:rates}(a); the normalized reconnection rate, $\alpha$, is plotted against the normalized system size, $L/d_{i}$, where $d_i$ is the ion skin depth, $d_i=c/\omega_{pi}$. Similar to other analyses of simulated reconnection \cite{stanier2015,sharma2019}, the rate increases as the normalized system size decreases. This is consistent with the transition toward electron-only reconnection, where the ions become less coupled to the field lines on the scale of the reconnection region, allowing reconnection to proceed without being constrained by the inertia of the ion fluid\cite{phan:2018,liu:2017}. The full analysis in Ref.~\onlinecite{olson:2021} also showed that the effective reconnection rate is constant regardless of the applied current drive $\left(\dot{I}\right)$; this is due to the interplay between particle density and magnetic field strength upstream of the layer. Even if the layer is forced down with a stronger drive, producing a larger $E_{rec}$ value, the shock structure and flux pileup will develop in such a way to produce a similar increase in the product of $B_{red}$ and $V_{a,hyb}$, resulting in a constant value for the scaled rate $\alpha$.

The TREX experiment is typically operated between two different density settings with three different ion species (hydrogen, deuterium, and helium), yielding six experimental points shown in Fig.~\ref{fig:rates}(a). To compare these results to simulated TREX setups in VPIC, we instead vary the value of initial plasma density. Examples of applied initial density profiles are shown in Fig.~\ref{fig:rates}(b). Our standard density profile vs $R$ is shown in red and labeled as $n$. To reach a range of values for the scaled system size, this standard density was varied up and down by a single factor multiplying the entire profile; for example, a profile of twice the standard density ($2n$) is shown in yellow, while another profile of half the standard density ($0.5n$) is shown in blue. Note that the profiles shown here have had their density gradient decreased from the actual profiles used in the simulations for the sake of clarity. The real density profiles are based on measurements in the experiment and show a much stronger dependence on the value of the $R$ coordinate.

Simulations were run for density values as low as $n/5$ and as high as $4n$, where $n$ represents our standard density. Three runs of each density value with different random initialization seeds were compared to reduce the chances of an anomalous result skewing the conclusions. Within each individual run, multiple points in time corresponding to the reconnection region being in the range of $R$ values most readily measured by the experiment ($R=0.35-0.45~\si{\meter}$) were selected and data from each of these points in time were sampled in regions $p$ and $r$. This data was then used to calculate the reconnection rate (following Eq.~\ref{eq:rate}-\ref{eq:vah}) and the local ion skin depth. The results from all these time points from each of the three repeated simulations of a given initial density setup were averaged together to produce the green points in Fig.~\ref{fig:rates}(a); the error bars represent the uncertainty estimate obtained by propagating the standard deviation of the distribution of selected density and magnetic fields through the rate equations. In general, these points follow the same trend as the experimental points (yellow), with increasing rate as the system size decreases.

One point of interest in Fig.~\ref{fig:rates}(a) is the dip in the green simulation rate results localized around $L/d_i \approx 2.2$. This feature is a real aspect of the data trend, tied to subtle Alfv\'{e}nic wave dynamics related to our cylindrical reconnection drive scenario. When the drive current begins to ramp up, the pressure balance of the initial configuration is suddenly violated, causing an Alfv\'{e}nic perturbation to propagate from beneath the drive coils downward toward $R=0~\si{\meter}$. Although this wavefront is propagating down, the wave itself corresponds to a radially outward expansion of the plasma. In the standard scenario with density scale $n$ (Fig.~\ref{fig:rates}(d) and (h)) this expansion persists throughout the reconnection layer formation and inward propagation, allowing the reconnection dynamics to adjust in a manner that keeps the effective rate consistent with expectation, as described earlier and in Ref.~\onlinecite{olson:2021}. However, in the lower density scenario ($n/3$, Fig.~\ref{fig:rates}(e) and (i)) the initial wavefront travels inward and then reflects off of the low $R$ boundary while the reconnection layer is still evolving. On the tailing side of the reflected wave-front the plasma expansion is significantly reduced, corresponding to a transient reduction in the drive as the front reaches the reconnection layer. The upstream conditions of the reconnection layer cannot instantly adjust to these effects, resulting in normalized reconnection drives that can be either enhanced or reduced. 

So far, this feature has not been clearly observed in the experiment, as it exists in a parameter regime that is not reachable in TREX. TREX has reached values of $L/d_i \approx 2.2$, but this was done with helium and deuterium plasmas rather than by going to lower density values. Furthermore, the effect of this Alfv\'{e}nic feature may be influenced by our simulation's reflecting boundary condition at low $R$, which approximates TREX's behaviour but may not be exactly analogous. This dip would not be expected in scenarios without some manner of reflection along one of the domain boundaries. Similarly, the results in Fig.~\ref{fig:rates}(c) show variation relative to drive current ramp intensity due to the above feature's effect in extinguishing the upstream inflow. This feature only appears in the simulations, causing them to diverge from the results of Ref.~\onlinecite{olson:2021} that showed experimentally that the drive intensity does not affect the scaled rate.

\section{\label{sec:conclusions}Conclusions}
Experiments conducted in the Terrestrial Reconnection EXperiment (TREX) over a range of different scaled system sizes showed a range of reconnection rates which increased as the system size decreased. As part of an ongoing effort to model the TREX experimental setup in a particle-in-cell simulation, VPIC was used to replicate TREX runs at a range of densities, many of which were outside TREX's normal operating parameters. Within the range of the TREX parameters, the numerical simulations confirmed the experimentally observed rates of reconnection, weakly dependent on the normalized size of the experiment with higher rates at smaller system size indicative of the transition toward electron-only reconnection. Additionally, the high detail of simulation data allows the full pressure balance equation across the reconnection shock front to be calculated and pressure balance to be confirmed. This calculation also allowed us to verify the accuracy of some of the assumptions that were needed in TREX's experimental pressure balance calculation. Together with previous results\cite{greess:2021}, these conclusions continue to verify VPIC's ability to accurately capture the full shock formation and reconnection dynamics observed in TREX.

\begin{acknowledgments}
We gratefully acknowledge DOE funds DE-SC0019153, DE-SC0013032, and DE-SC0010463 and NASA fund 80NSSC18K1231 for support of the TREX experiment. In addition, the experimental work is supported through the WiPPL User Facility under DOE fund DE-SC0018266. Simulation work was supported by the DOE Basic Plasma Science program and by a fellowship from the Center for Space and Earth Science (CSES) at LANL. CSES is funded by LANL’s Laboratory Directed Research and Development (LDRD) program under project number 20180475DR. This work used resources provided by the Los Alamos National Laboratory Institutional Computing Program, which is supported by the U.S. Department of Energy National Nuclear Security Administration under Contract No. 89233218CNA000001.
\newline
\end{acknowledgments}

\section*{Data Availability Statement}
The data that support the findings of this study are available from the corresponding author upon reasonable request.

\bibliography{references}

\end{document}